\documentclass[twocolumn, amsmath, amssymb, aps, prx, superscriptaddress]{revtex4-1}

\usepackage{graphicx}   
\usepackage{dcolumn}  
\usepackage{bm}          
\usepackage{xcolor}    

\begin{document}
\title{Viscoelastic and poroelastic relaxations of soft solid surfaces}
\author{Qin Xu}
\email[]{qinxu@ust.hk}
\affiliation{Laboratory of Soft and Living Materials, ETH Zurich, 8093 Zurich, Switzerland}

\affiliation{Department of Physics, The Hong Kong University of Science and Technology, Hong Kong, China}

\affiliation{HKUST Shenzhen Research Institute, Shenzhen, China 518057}

\author{Lawrence~A. Wilen}\affiliation{School of Engineering and Applied Science, Yale University, New Haven, Connecticut, USA 06511}
\author{Katharine~E. Jensen}\affiliation{Department of Physics, Williams College, Williamstown, Massachusetts, USA 01267}
\author{Robert~W. Style}\affiliation{Laboratory of Soft and Living Materials, ETH Zurich, 8093 Zurich, Switzerland}
\author{Eric~R. Dufresne}
\email{eric.dufresne@mat.ethz.ch}
\affiliation{Laboratory of Soft and Living Materials, ETH Zurich, 8093 Zurich, Switzerland}

\begin{abstract}
Understanding surface mechanics of soft solids, such as soft polymeric gels, is crucial in many engineering processes, such as dynamic wetting and adhesive failure. In these situations, a combination of capillary and elastic forces drives the motion, which is balanced by dissipative mechanisms to determine the rate. While  shear rheology (\emph{i.e.} viscoelasticity) has long been assumed to dominate the dissipation, recent works have suggested that compressibility effects (\emph{i.e.} poroelasticity) could play roles in swollen networks. We use fast interferometric imaging to quantify the relaxation of surface deformations due to a displaced contact line. By systematically measuring the  profiles at different time and length scales, we experimentally observe a crossover from  viscoelastic to poroelastic surface relaxations. 
\end{abstract}
\maketitle

The mechanics of soft solids has recently drawn great attention for their potential use in various applications, such as design of bio-compatible materials \cite{RevModPhys.85.1327, Gonzalez_Rodriguez2012, AM_Zhao2015, Zhigang2019}, cell patterning \cite{Whitesides2001,Matsumoto2007}, machining of soft robotics \cite{whitesides2014}, and fabrication of microfluidic devices \cite{Khare2009}. These applications rely upon the contact of soft solids with other materials where their wetting and adhesive properties play essential roles. 

Soft solids can  deform strongly at contact lines, where they meet the interfaces of two other phases \cite{Dufresne_annual_review2017}.
For example, a droplet's liquid-vapor interface creates a ridge on a soft solid where the bulk elastic stresses are balanced by the solid surface tension \cite{Andreotti2020}.  As a droplet slides on a soft solid, the displacement of the ridge is found to significantly slow down its movement.
First observed in \cite{Carre1996}, this phenomena was  dubbed `viscoelastic braking' because the substrate's viscoelasticity was presumed to be the underlying dissipation \cite{Long1996, Karpitschka2015,Zhao2018,Karpitschka2018,Grocum2020}. 
While shear rheology dominates the response of most soft materials, this can break down for polymeric gels. 
A gel is an elastic network swollen by a fluid. 
Even when the elastic network is easily compressed, the solvent avoids compression by flowing through the network. This causes local changes to the relative concentration of solvent and the network. The coupling of fluid flow to the deformation of an elastic network is called \emph{poroelasticity}. The theory of poroelasticity was originally developed for geological applications \cite{Bio1941,DETOURNAY1993113}, but has recently been applied to describe the deformation of hydrogels \cite{Hu2010, Cai2010, Hu2012}.

Recent experiments suggest that the flow of solvent through a gel's elastic network could impact wetting  and adhesion.
Zhao \emph{et al} found a slow, approximately logarithmic, relaxation of a wetting ridge upon removal of a droplet, and introduced a poroelastic model for the decay \cite{Zhao2017_soft}.
Berman \emph{et al} measured the relaxation of an adhesive contact and made a scaling argument for the importance of solvent flow in determining the dynamics of relaxation \cite{Kate2019}. 
These arguments are supported by separate experiments which have quantified how contact lines can extract solvent from the bulk \cite{Jensen2015},  coating the droplet and reducing its surface tension \cite{Hourlier-Fargette2018}. 

Despite  evidence supporting the relevance of both viscoelastic and poroelastic effects to the surface relaxations of soft gels, previous works have only focused on one relaxation mechanism at the exclusion of the other and there is no clear framework for evaluating which of these mechanisms will dominate in a particular situation. The ambiguity is caused by the difficulty in precisely measuring surface dynamics over a broad range of time and length scales. 

In this Letter, we apply direct interferometric imaging to measure the relaxation of a wetting ridge after a sudden displacement of the contact line. 
We observe contributions from both viscoelastic and poroelastic dissipation in the same dewetting process. 
For relatively large droplets and long timescales,  dynamics are dominated by the substrate's poroelastic response.
On the other hand, viscoelasticity dominates for small droplets or short timescales.

\begin{figure}[h]
\centering
\includegraphics[width=82mm]{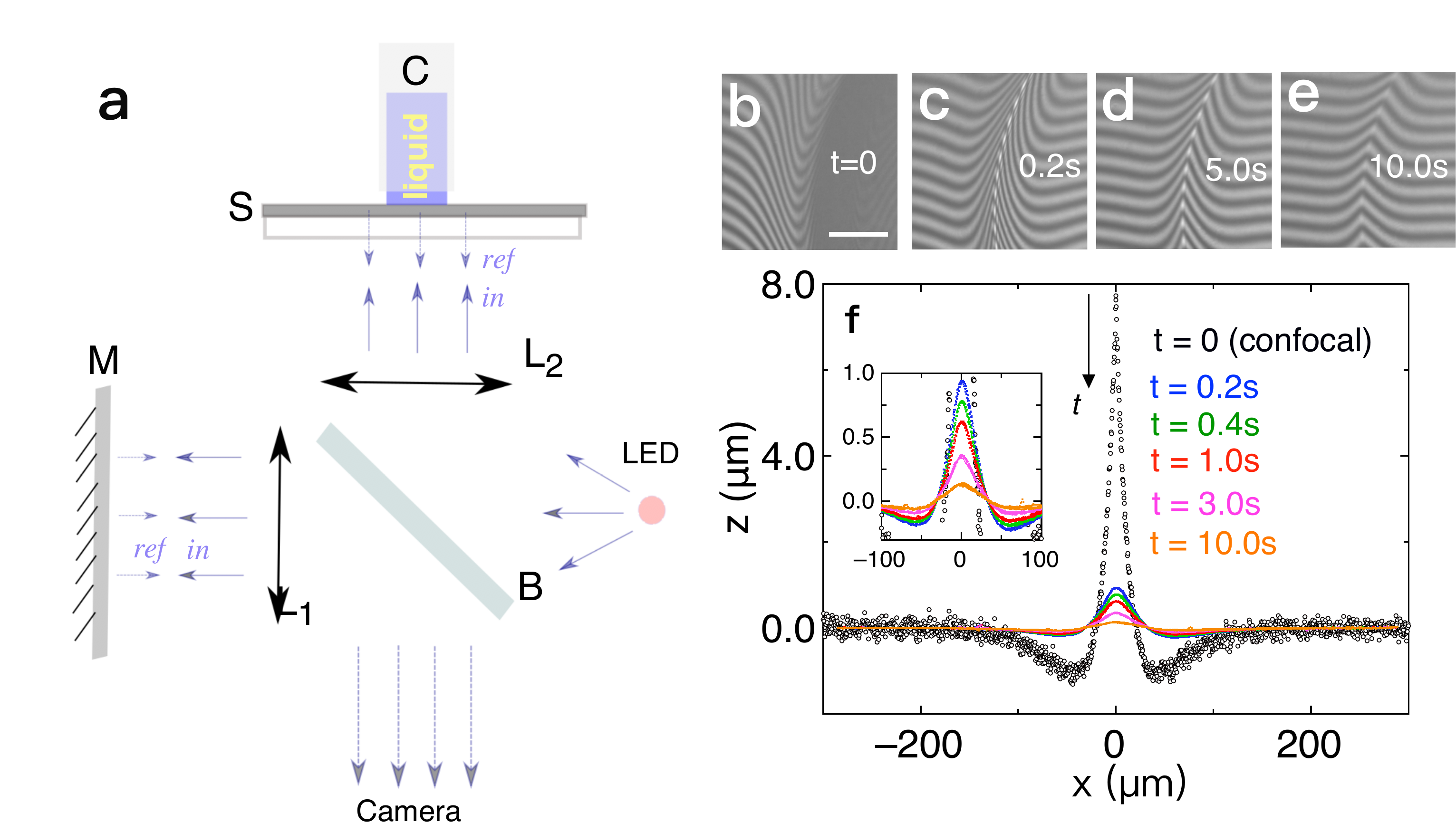}
\caption{\emph{Interferometric imaging}. {\bf (a)} Schematic showing major components of the imaging system, including a monochromatic LED light source ($\lambda\pm \Delta\lambda = 643\pm 22$ nm), the  sample stage (S) with a capillary tube (C) to deposit the liquid, a reference mirror (M), two focusing lenses (L1 \& L2) and a beam splitter (B).  {\bf (b-e)} Series of unprocesssed interference images from the gel surface after forced dewetting at $t=0$.  Scale bar: 100 $\mu$m. {\bf (f)} Surface profile just before dewetting ($t<0$)  obtained by confocal microscopy (black dots) and surface profiles after dewetting ( $t=$ 0.2, 0.4, 1.0, 3.0 and 10.0s) reconstructed from interferograms (colored dots). Inset: zoom-in $t>0$ data.}

\label{fig:relaxation}
\end{figure}

Our experimental approach is  illustrated schematically in Fig. \ref{fig:relaxation}(a). To image the surface deformation of the substrate, we designed a Linnik interference imaging system \cite{Dubois2002}. This technique quantifies interface topography with a vertical resolution about 10 nm and a temporal resolution limited by the speed of the camera, here about 20 ms. These specifications are superior to confocal microscopy and other methods that have been used to image wetting ridges, but come with some costs. Most importantly, interference microscopy cannot image height gradients bigger than about 35\% \cite{SI}. We prepare silicone gel (Dow Corning CY52-276) substrates by spin-coating the curing silicone mixture on a microscope slide at 800 rpm for a minute. After the gel is fully cured at 40$^\circ$C, it forms a 65 $\mu$m thick smooth substrate with a Young's modulus of 3.8 kPa.

To perform dewetting experiments, we deposit liquid on the surface of soft gels by a capillary tube. At the contact line, the droplet's surface tension pulls on the substrate, forming a microscopic wetting ridge \cite{Carre1996, pericet2008}. In equilibrium, the height of the wetting ridge, $h_0$, scales as $ \gamma_l\sin\theta/E$. Here, $\gamma_l$ is the liquid-vapor surface tension and $\theta$ is the macroscopic contact angle \cite{Style2012}. For water on a $\mathcal{O}$(kPa) substrate, the height of the wetting ridge is a few microns and the contact angle is around 91$^\circ$ \cite{Xu2017}. After a certain resting time $t_{res}$, we suddenly 
pull away the droplet to remove the liquid-vapor interface. To avoid the effects of pinning between liquid and soft gels, the retraction speed of contact line is always kept larger than 1 mm/s. The relaxation of the wetting ridge was imaged by the camera from below. 

Unprocessed interferograms, shown in Figs. \ref{fig:relaxation}(b-e),  show the relaxation of the interface upon removal of a 0.45 mm radius water droplet after a residence time of $t_{res}=10^3$ s. 
The dark region on the right of panel (b) corresponds to reflection from the solid-liquid interface, which has a smaller index-mismatch than the solid-vapor interface on the left. 
There, clearly resolved fringes indicate the surface topography of the solid-vapor interface. 
Far from the wetting ridge, they are parallel and evenly spaced due to the gentle tilt of the otherwise flat interface.
When the droplet is removed, both sides of the wetting ridge have equal contrast and the decay is directly observed, as shown in panels (c-e) . 
Digital processing of these fringes enables precise quantification of the ridge profile and its decay \cite{Takeda82}, shown by the colored data points in Fig. \ref{fig:relaxation}(f). 
Note that the interference profiles start at $\sim 0.2~\textrm{s}$ after removal of the droplet. 
At earlier times, the center of the full profile is too steep ($>35\%$) to show interference fringes. For comparison, we superimpose the steady wetting profile of a similar-sized droplet on the same substrate quantified by confocal microscopy, shown by the black data points in Fig. \ref{fig:relaxation}(f). 
This suggests that the ridge quickly retracts from $7.8$ $\mu$m to $0.9$ $\mu$m in the early stage ( $\sim 10^{-1}$ s), followed by a slow relaxation process in a period of $\sim 10^1$ s. By varying the droplet residence time $t_{res}$ before removing the contact line, we found that the surface relaxation becomes consistent when $t_{res} > 10$ mins \cite{SI}. Thus, for all the experiments shown in the rest of the paper, we keep $t_{res}$ to be at least 10 mins to ensure surface relaxations from a fully developed wetting ridge.

\begin{figure}[t]
\centering
\includegraphics[width=80mm]{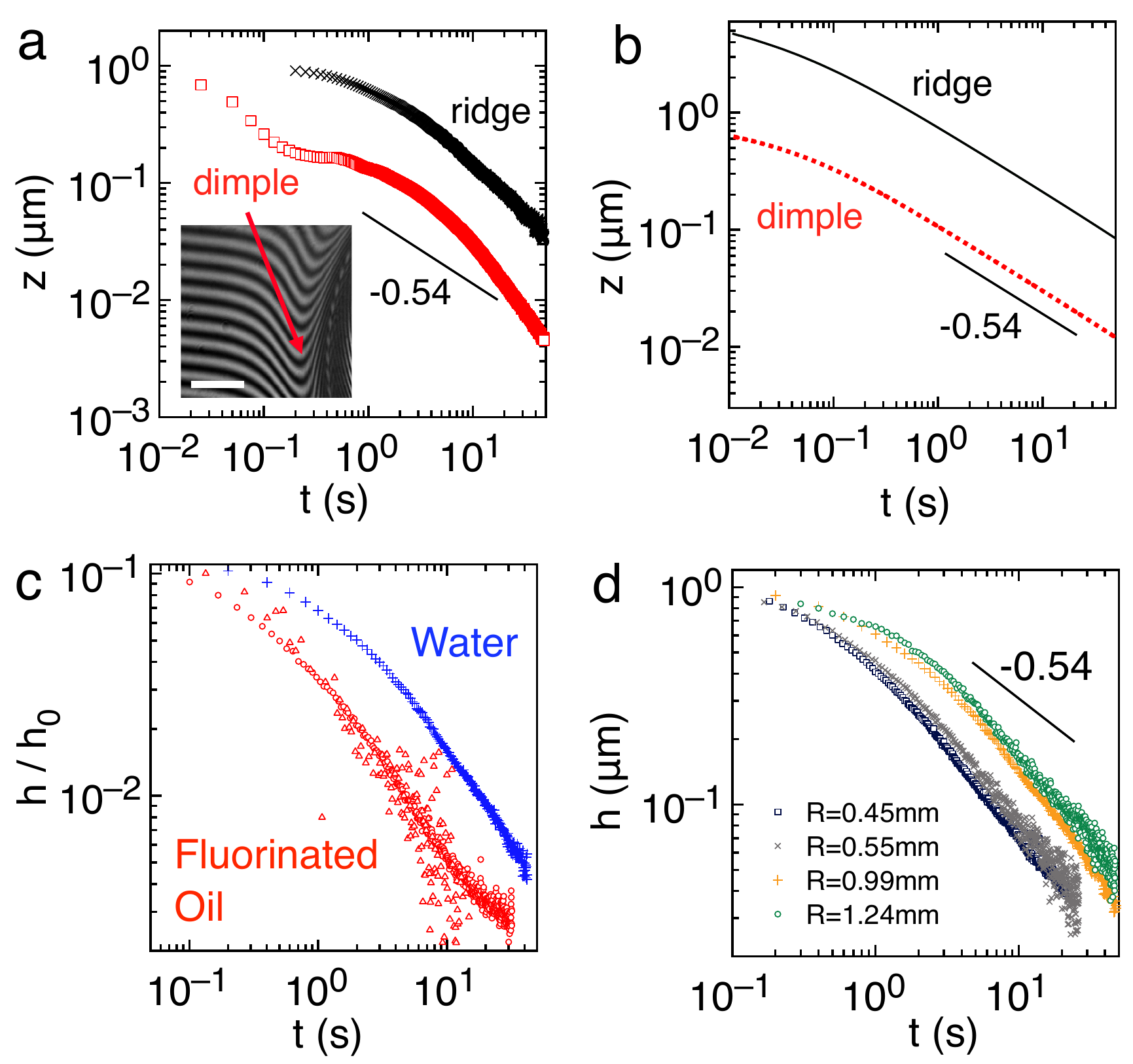}
\caption{ \emph{Variation of relaxation dynamics with waiting time and deformation sizes.} {\bf (a)} Relaxation plots of the ridge height and dimple depth measured in experiments for $1.1$ mm radius droplet. Inset: Interference image showing the `dimple.' Scale bar: 100 $\mu m$. {\bf (b)} Theoretical calculation results of the ridge and dimple relaxations based on power-law rheology. {\bf (c)} Measured $h(t)/h_0$  for different wetting liquids (water and fluorinated oil).{\bf (d)} Measured ridge height, $h$ \emph{versus} time, $t$, for various water droplet radii $R$. }

\label{fig:time-and-scale-dependent}
\end{figure}

We compare the experimental relaxation of the surface  with the theoretical predictions for a purely viscoelastic relaxation.
The height of the ridge and the depth of the dimples are plotted as black crosses and red-squares respectively in Fig. \ref{fig:time-and-scale-dependent}(a).  
The dimples are the small indentations on either side of the wetting ridge, seen clearly in Fig. \ref{fig:relaxation}f,  which arise from a combination of near-incompressibility of the gel and finite thickness of the substrate \cite{Style2013_PRL}.
Because the slope of the surface is small near the dimples, we can reliably measure their depth at much earlier times than we can the height of the wetting ridge.

To calculate the expected viscolelastic response, we need to measure the viscoelastic spectrum, shown in Fig. S5.
We find that the complex modulus is well described by the form $G^\star(\omega) = G_0 (1+ (i \omega\tau_c)^n)$, with an index $n =0.54$ and intrinsic time scale $\tau_c\approx 0.11$s.
 Following the theoretical model presented in Ref. \cite{Karpitschka2015}, we numerically calculated the expected time-dependence of the ridge height and dimple depth, shown respectively as solid black and dashed red lines in Fig. \ref{fig:time-and-scale-dependent}(b).  This calculation is based not only on the measured rheology, but also on the substrate thickness and the measured surface stresses of water after being in contact with the silicone substrate (46 mN/m) and solid gel (31 mN/m).  For both the dimple and ridge, the viscoelastic model predicts that the height decays as $\sim t^{-n}$ for $t>\tau_c$.

The predictions of the viscoelastic model do not capture the observed surface relaxation.  At late times, neither the ridge nor the dimple show the expected power law relaxation. At short times, the dimple shows a fast relaxation followed by a plateau.  This feature is missing from the prediction of the viscoelastic model with power-law rheology.  On the other hand, this deviation occurs at time-scales where the rheological data is sparse and subject to systematic errors due to tool inertia. Therefore, we will now focus our attention on the long-time relaxation.

Further limitations of the viscoelastic model are  suggested by the dependence of the relaxation on the height of the wetting ridge. To explore the effect of ridge height, we compare the ridge relaxation of fluorinated oil droplets with water droplets of the same radii ($R =1$ mm) in Fig.\ref{fig:time-and-scale-dependent}(c).   If  viscoelasticity were the dominant dissipation mechanism, then normalizing  the ridge height by its initial value, $h_0$, would be sufficient to collapse the height relaxation curves in Fig.\ref{fig:time-and-scale-dependent}(c), as described in \cite{Karpitschka2015}.  Using confocal microscopy, we measured  initial ridge  heights ($h_0$) of 7.8  $\mathrm{\mu m}$ for water (Fig. \ref{fig:relaxation}(f)) and 2.1 $\mathrm{\mu m}$ for fluorinated oil (Fig. S4 in \cite{SI}). 
However, this normalization does not collapse the data: for the same droplet radius, shorter ridges decay even faster than expected.

Finally, the surface relaxations are found to vary with droplet radius, $R$. As shown by the blue symbols (crosses, circles, stars, and squares) in Fig. \ref{fig:time-and-scale-dependent}(d), the decay time of ridge height increases significantly as the radius of water droplet increases from $R = 0.45$ to $1.24$ mm. 
The existing viscolastic model \cite{Karpitschka2015} was formulated for a straight contact line, and therefore does not account for droplet size dependence. 
However, it is known that the height of the wetting ridge increases slowly with the droplet radius, even when its radius is much bigger than elastocapillary length or the substrate thickness \cite{Style2012}.    
For droplet radii ranging from  0.45 mm to 1.24 mm, the ridge height increases by less than $5$\% \cite{Style2013_PRL}.  
Normalizing the relaxation curves by their initial height is therefore insufficient to collapse the data, whose timescales are shifted by almost a factor of two.   Considering  the collective evidence presented in Fig. \ref{fig:time-and-scale-dependent}, we conclude that  viscoelastic dissipation is insufficient to  describe the relaxation of the wetting ridge.

\begin{figure}[t]
\includegraphics[width=85mm]{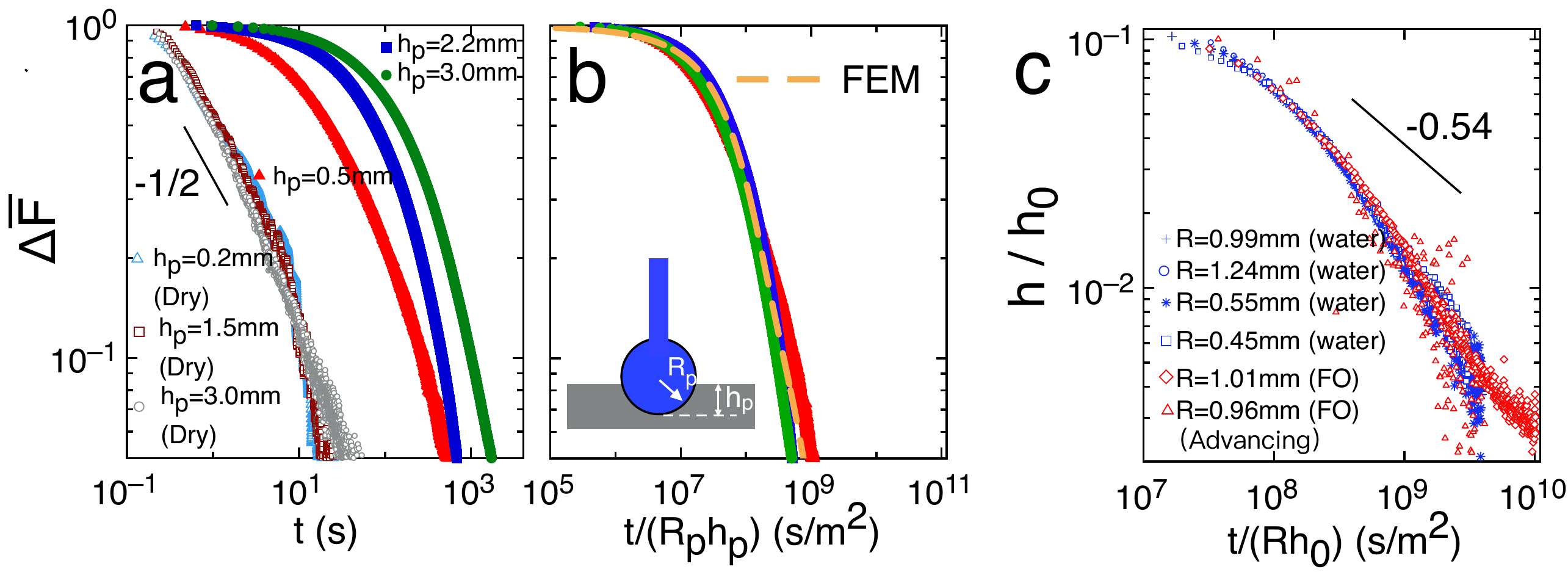}
\caption{ \emph{Quantifying poroelastic response}.  {\bf (a)} Relaxation of normal force after indentation. Measured normalized force, $\Delta \bar{F}$, \emph{versus} time, $t$, for different indentation depths, $h_p$, on soft gels (solid dots) and solvent-extracted (`dry') networks (hollow dots). {\bf (b)} Force relaxation data of the gels from (a) collapsed by normalizing  time with  $R_p h_p$. Inset:  schematic of the indentation experiments. {\bf (c)} Surface profile data from Fig. \ref{fig:relaxation}(b) collapsed by normalizing the height by $h_0$ and time by $Rh_0$.}

\label{fig:poroelasticity}
\end{figure}

Poroelastic relaxations, by contrast, are distributed and scale-dependent \cite{Hu2010}. Stresses in a porous elastic solid drive flow of fluid through its pores.  For  deformations observed at  length scales much larger than the pores, compositional differences in a porous medium evolve according to the familiar diffusion equation with an effective diffusion coefficient, 
\begin{equation}
       D^\star=\frac{2G_0 k(1-\nu)}{\mu(1-2\nu)}.
\label{eq:diffusion} 
\end{equation}
Here, the elastic network is characterized by its shear modulus, $G_0$, Poisson ratio, $\nu$, and permeability, $k$. The fluid is characterised by its shear viscosity, $\mu$.  From inspection of Eq. \ref{eq:diffusion}, it is clear that poroelastic diffusion can only be significant for compressible networks  ($\nu<1/2$). Since poroelastic relaxation is diffusive, its characteristic time should scale as $\tau_p\sim L_D^2/D^\star$. Here, $L_D$ is a characteristic length scale set by the macroscopic geometry of the deformation.

Suo and collaborators \cite{Hu2011} have recently proposed standard mechanical tests to measure poroelastic properties of soft gels by compressing the elastic network suddenly at t=0 and measuring the relaxation of normal force $F(t)$ for different indentation depths. The relaxation dependence on initial deformation can help to measure the effective diffusivity $D^\star$ \cite{SI}. We applied this method to independently assess the poroelastic properties of our soft gels, indenting with a steel sphere of radius $R_p=9.5$ mm to a small distance ($h_p\ll R_p$). 
As a control experiment, we also repeated the same tests on solvent-extracted `dry networks' from the same gels that is obtained via a standard toluene extraction procedure \cite{Hourlier-Fargette2018,SI}. The normalized force relaxation curves, $\Delta \bar{F}(t)= (F(t)-F(\infty))/(F(0)-F(\infty))$, are shown for both the gel and  dry network in Fig. \ref{fig:poroelasticity}(a). 
While $\Delta \bar{F}(t)$ of the original soft gels, indicated by the solid dots, varies significantly with the indentation depth $h_p$, the dry networks show no dependence on depth, as shown by the hollow points in panel (a). 
The observed scaling of force relaxation in the dry network, $\Delta \bar{F} \sim t^{-n}$ with $n\approx 1/2$, is consistent with its power-law rheology.
To measure the poroelastic diffusivity $D^\star$ of the gels, we apply results by Hu \emph{et al} \cite{Hu2010} who showed that the diffusive length scale for a spherical indenter is given by $L_D=\sqrt{R_p h_p}$. 
Normalizing the timescale of the relaxation by $L_D^2$, 
as displayed in Fig. \ref{fig:poroelasticity}(b), the force relaxation curves collapse and match the master curve predicted by FEM simulation in \cite{Hu2010} for a poroelastic diffusion coefficient of $D^\star=6.1\times 10^{-9}$ m$^2$/s.

In an indentation experiment, the porelastic relaxation time is set by the indenter radius, $R_p$, and indentation depth, $h_p$. 
What length scales determine the relaxation time in our dewetting experiments, where there is no apparent indentation, only the formation of a wetting ridge?
In the absence of a theoretical prediction, we take an empirical approach.
It is clear from Figs. \ref{fig:time-and-scale-dependent}(c) and (d) that the droplet radius and initial ridge height are relevant factors. 
 We found  that a diffusion length $L_D\sim\sqrt{Rh_0}$ nicely collapses the height relaxation data for all droplet sizes and compositions in Fig. \ref{fig:poroelasticity}(c). 
 
The decay of the master curve in Fig. \ref{fig:poroelasticity}(c) suggests a normalized diffusion time $t/(Rh_0)\sim 10^8$ s/m$^2$, and therefore a poroelastic diffusion coefficient of order $D^\star\approx10^{-8}~\mathrm{m^2/s}$. 
 This value is consistent with the measured poroelastic diffusion coefficient by stress-relaxation. Precise quantification of the ridge relaxations for dry networks was not possible because solvent extraction leads to strong wrinkling effects on the surface and possible stress localization could affect the behaviors. Qualitative observations of interferograms (shown in the Supplement Fig. S7 \cite{SI}), however, suggest a timescale of $\sim 10^{-1}$s, comparable to the viscoelastic timescale $\tau_c$, once the free solvent was fully extracted. Therefore, while the long-time relaxation is governed by poroelastic flow, the short time relaxation  (e.g., the dimple profile in Fig.\ref{fig:time-and-scale-dependent}b) is possibly characterized by the viscoelastic nature of the network.

\begin{figure}[t]
\centering
\includegraphics[width = 70 mm]{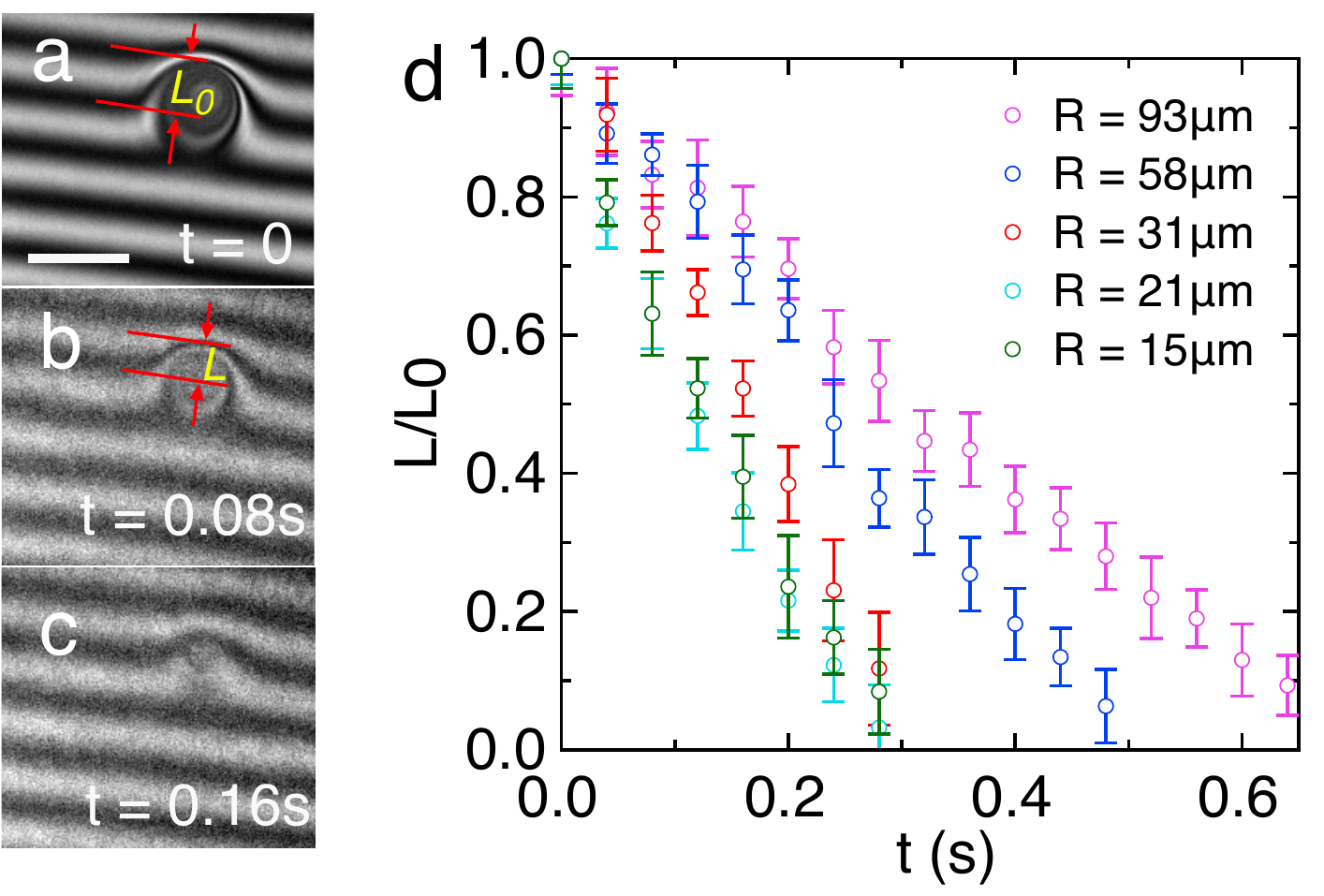}
\caption{\emph{Dominance of viscoelasticity at small length scales and short times}. {\bf (a)-(c)} Series of raw interference images showing the relaxation of the substrate after a 21 $\mu$m radius glycerol droplet has been removed by flooding. We define $L$ as the maximum deformation of the fringes due to the presence of liquid droplets, and the initial deformation is indicated by $L_0$. Scale bar: 50 $\mu$m. {\bf (d)} Plots of $L/L_0$  vs. $t$ for different sizes of droplets on soft gels. The error bars  indicate the uncertainty in determining $L$ from the image analysis.}
\label{fig:small}
\end{figure}

We have found that the poroelastic timescale governing the long time relaxation of a wetting ridge, $\tau_p$, scales as  $Rh_0/D^\star$. In these experiments with millimeter-scale droplets, poroelastic relaxation time $\tau_p$ is  larger than $\tau_c$,  and therefore is rate-limiting.
We expect a cross-over to a regime where viscoelastic response is rate-limiting for small droplets, $R < D^\star \tau_c / h_0$. 
For water droplets, our data suggest that this cross-over should occur for radii around  $50~\mathrm{\mu m}$.
To work with small droplets, however, we need to make a few modifications of our technique.
First, we use glycerol, since it has a lower vapor pressure, but similar surface tension to water.
Second,  we disperse small droplets across the gel surfaces with an atomizer (Misto). 
Finally, we remove the liquid-air interface by flooding the surface with glycerol.
 Figures \ref{fig:small}(a-c) exhibits a series of interference images immediately after a $R=21$ $\mu$m droplet has been removed by flooding.  Note that the topography of such a submerged interface is not accessible by conventional first-surface interferometry, but requires the Linnik method.
Since the droplet is too small to obtain the precise wetting profile from the image, we simply measure the maximum deflection of the interference fringe near the droplet, which we will call $L$.
Since this value depends not just on the surface profile, but the droplet's position relative to the fringe, it cannot provide us with absolute information about the height of the ridge.
However, its time evolution still contains useful information about the dynamics of relaxation.
Thus, we consider $L(t)/L(0)$, shown in Fig. \ref{fig:small}(d).
To separate viscoelastic and poroleastic effects, we varied the droplet radii  $R$ from $15$ to $93$ $\mu$m.
For droplets with radii less than $50~\mathrm{\mu m}$, the relaxation is size independent, with a characteristic time scale in the order of $\sim 10^{-1}$s, comparable to $\tau_c$.
Thus, we conclude that, with our temporal resolution, viscoelastic dissipation dominates as $\tau_p < \tau_c$.  
Note that a direct comparison with a viscoelastic theory is not yet possible for this case, because the current theory is formulated for large droplets, whose radius is much larger than the elastocapillary length \cite{Karpitschka2015}.

Our observation shows signatures of both viscoelastic and poroelastic relaxations on soft gel surfaces. 
The dominant mechanism is determined by the experimental timescale, the viscoelastic characteristic time, $\tau_c$, and the poroelastic relaxation time, ${L_D^2}/{D^\star}$. 
In our dewetting experiments, the full relaxation of the wetting ridge required solvent transport over a length scale  $(Rh_0)^{1/2}$.
Further theoretical analysis is needed to understand the origin of this scaling.
Intriguingly, the slow and steady sliding of wetting ridges on the same material has been shown to be limited by the viscoelastic response of the substrate \cite{Karpitschka2015}.
We see no contradiction with the current results, as a slow sliding wetting ridge involves no large-scale displacement of material.
In that case, we expect $L_D \lesssim h_0$, and correspondingly very short poroelastic relaxation times ($\tau_p \ll \tau_c$). 

A closer look at the microscopic origins of the poroelastic diffusion coefficient is further  warranted.
A simple model combining Darcy flow and rubber elasticity suggests  $D^\star \sim G_0^{1/3}(k_B T)^{2/3}\mu^{-1}$.
While this works very well for hydrogels \cite{hu_chen2011}, it underestimates the diffusion coefficient of our soft silicone gels by $60$ times.
Many factors could contribute to this discrepancy.  
However, we suspect non-affine deformation of the pore-space or a breakdown of Darcy's law due to a lack of separation of the structural length scales of the solvent and network.
The need for further analysis of the poroelastic response of polymer networks with polymeric solvents is further highlighted by the drastically different responses from nearly identical silicone systems.
We completed the full battery of tests described in the main body of this paper on a second soft silicone gel (Gelest DMS-V31) with nearly identical shear rheology and a similar fraction of uncrosslinked chains.  
Despite these similarities, we saw no evidence of poroelastic response in dewetting experiments (Figs. S5 and S6 in \cite{SI}). 
All of the responses were consistent with a purely viscoelastic response, and a much higher poroelastic diffusion coefficient \cite{berman2019}.
Thus, shear rheology alone is a poor predictor of the dynamics of relaxation on soft gels, and further efforts are required to determine the microscopic origins of the poroelastic response of networks with free chains.

We acknowledge Dr. Julien Dervaux, Prof. Anand Jagota, Dr. Stefanie Heyden and Prof. Michael Loewenberg for useful discussions. 

\bibliography{reference_qx}
\end{document}